\documentclass{article}

\usepackage{graphics}
 
\begin{document}

\newtheorem{thm1}{Theorem}
\newtheorem{thm2}[thm1]{Theorem}

\def\oti{{\otimes}}
\def\bra#1{{\langle #1 |  }}
\def\lb{ \left[ }
\def\rb{ \right]  }
\def\tilde{\widetilde}
\def\bar{\overline}
\def\*{\star} 

\def\({\left(}          \def\BL{\Bigr(}
\def\){\right)}         \def\BR{\Bigr)}
        \def\BBL{\lb}
        \def\BBR{\rb}
%

\def\tr{ {\rm{Tr }}\,}
\def\pr{ {\rm{Pr }}}
\def\E{{\mathbf{E} }}
\def\1{{\mathbf{1} }}

\def\bb{{\bar{b} }}
\def\ab{{\bar{a} }}
\def\zb{{\bar{z} }}
\def\zbar{{\bar{z} }}
\def\frac#1#2{{#1 \over #2}}
\def\inv#1{{1 \over #1}}
\def\half{{1 \over 2}}
\def\d{\partial}
\def\der#1{{\partial \over \partial #1}}
\def\dd#1#2{{\partial #1 \over \partial #2}}
\def\vev#1{\langle #1 \rangle}
\def\ket#1{ | #1 \rangle}
\def\rvac{\hbox{$\vert 0\rangle$}}
\def\lvac{\hbox{$\langle 0 \vert $}}
\def\2pi{\hbox{$2\pi i$}}
\def\e#1{{\rm e}^{^{\textstyle #1}}}
\def\grad#1{\,\nabla\!_{{#1}}\,}
\def\dsl{\raise.15ex\hbox{/}\kern-.57em\partial}
\def\Dsl{\,\raise.15ex\hbox{/}\mkern-.13.5mu D}
\def\b#1{\mathbf{#1}}
%
%
\def\th{\theta}         \def\Th{\Theta}
\def\ga{\gamma}         \def\Ga{\Gamma}
\def\be{\beta}
\def\al{\alpha}
\def\ep{\epsilon}
\def\vep{\varepsilon}
\def\la{\lambda}        \def\La{\Lambda}
\def\de{\delta}         \def\De{\Delta}
\def\om{\omega}         \def\Om{\Omega}
\def\sig{\sigma}        \def\Sig{\Sigma}
\def\vphi{\varphi}
%
%
\def\CA{{\cal A}}       \def\CB{{\cal B}}       \def\CC{{\cal C}}
\def\CD{{\cal D}}       \def\CE{{\cal E}}       \def\CF{{\cal F}}
\def\CG{{\cal G}}       \def\CH{{\cal H}}       \def\CI{{\cal J}}
\def\CJ{{\cal J}}       \def\CK{{\cal K}}       \def\CL{{\cal L}}

\def\CM{{\cal M}}       \def\CN{{\cal N}}       \def\CO{{\cal O}}
\def\CP{{\cal P}}       \def\CQ{{\cal Q}}       \def\CR{{\cal R}}
\def\CS{{\cal S}}       \def\CT{{\cal T}}       \def\CU{{\cal U}}
\def\CV{{\cal V}}       \def\CW{{\cal W}}       \def\CX{{\cal X}}
\def\CY{{\cal Y}}       \def\CZ{{\cal Z}}

\def\rvac{\hbox{$\vert 0\rangle$}}
\def\lvac{\hbox{$\langle 0 \vert $}}
\def\comm#1#2{ \BBL\ #1\ ,\ #2 \BBR }
\def\2pi{\hbox{$2\pi i$}}
\def\e#1{{\rm e}^{^{\textstyle #1}}}
\def\grad#1{\,\nabla\!_{{#1}}\,}
\def\dsl{\raise.15ex\hbox{/}\kern-.57em\partial}
\def\Dsl{\,\raise.15ex\hbox{/}\mkern-.13.5mu D}
\def\beq{\begin {equation}}
\def\eeq{\end {equation}}
\def\to{\rightarrow}
\newcommand{\qed}{\rule{7pt}{7pt}}

\title{Classical data compression with quantum side information }

\author{I. Devetak\footnote{Electronic address: devetak@us.ibm.com} \\
\it{IBM T.J. Watson Research Center, Yorktown Heights, NY 10598, USA} \\
\\
 A. Winter\footnote{Electronic address: winter@cs.bris.ac.uk} \\
\it{Department of Computer Science, University of Bristol, Bristol BS8 1UB, U.K.}
 } 
  \date{\today} 
  \maketitle
  
\begin{abstract}
The problem of classical data compression 
when the decoder has quantum side information at his disposal 
is considered. This is a  quantum generalization of the classical 
Slepian-Wolf theorem. The optimal compression rate is found 
to be reduced from the Shannon entropy of the source by
the Holevo information between the source and side information.
\end{abstract}

Generalizing classical information theory to the quantum setting 
has had varying success depending on the type of problem considered. 
Quantum problems hitherto solved (in the asymptotic sense of Shannon theory)
may be divided into three classes. The first comprises pure bipartite 
entanglement manipulation, such as Schumacher compression \cite{nono} and entanglement 
concentration/dilution \cite{BBPS,lo, koren}. Their tractability is due to the 
formal similarities between a pair of perfectly correlated random 
variables and the Schmidt decomposition of bipartite quantum states.

The second, and largest, is the class of ``hybrid'' classical-quantum problems, 
where only a subset (usually of size one) of the terminals in the problem is quantum 
and the others are classical.
The simplest example is the Holevo-Schumacher-Westmoreland (HSW) theorem 
\cite{holevo}, which deals with the capacity of a 
classical $\rightarrow$ quantum channel (abbreviated 
$\{ c \rightarrow q \}$; see \cite{lock}).
This carries over to the multiterminal case involving many classical senders 
and one quantum receiver \cite{mac}. Then we have Winter's measurement compression
theorem \cite{winter}, and remote state preparation \cite{RSP, devberger, newRSP}. 
These two may be thought of as simulating quantum $\rightarrow$ classical 
($ \{ q \rightarrow c \}$) and $\{ c \rightarrow q \}$ channels, respectively. 
Another recent discovery has been quantum 
data compression with classical side-information available to both
the encoder and decoder \cite{HJW}, generalizing 
the rate-entropy curve of \cite{devberger} to arbitrary pure state ensembles.

The third class is that of fully quantum communication problems,
such as the entanglement-assisted capacity theorem \cite{eac} and
its reverse -- that of  simulating quantum channels in the presence of 
entanglement \cite{shor}. 
These rely on methods of $\{ c \rightarrow q \}$ channel coding
combined with super-dense coding \cite{dense}
and  $\{ q \rightarrow c \}$ channel simulation combined with
quantum teleportation \cite{tele}, respectively.
A recent addition to this class has been the long awaited 
proof of the channel capacity theorem \cite{coh, devetak}, 
which also relies on classical-quantum methods. 

The problem addressed here belongs to the second class and concerns classical 
data compression when the decoder has quantum side information at his disposal. 
We shall refer to it as the 
classical-quantum Slepian-Wolf (CQSW) problem in analogy to its classical 
counterpart \cite{slepian}. We begin by introducing the notion of a bipartite 
\emph{classical-quantum system}. The fully classical and
fully quantum analogues are familiar concepts. The former is
is embodied in a pair of correlated random variables $XY$, 
associated with the product set $\CX \times \CY$ and a probability 
distribution $p(x,y) = \pr\{ X = x, Y = y\}$ defined on $\CX \times \CY $.  
The latter is a bipartite quantum system $\CA \CB$,
associated with a product Hilbert space $\CH_\CA \otimes \CH_\CB$ 
and a density operator $\rho^{\CA \CB}$,
the ``quantum state'' of the system $\CA \CB$, defined on $\CH_\CA \otimes \CH_\CB$. 
The state of a classical-quantum system $X \CQ$ is now
described by an \emph{ensemble} $\CE = \{\rho_x, p(x) \}$,
with $p(x)$ defined on $\CX$ and the $\rho_x$ being density operators on
the Hilbert space $\CH_\CQ$ of $\CQ$. Thus, with probability $p(x)$
the classical index and quantum state take on values  $x$ and $\rho_x$, 
respectively. Such correlations may come about, 
for example, when Bob holds the purification of a state Alice is measuring.
Indeed, let Alice and Bob initially share the quantum state (in Schmidt polar form) 
$$
\ket{\Phi}_{\CA \CB} = \sum_i \sqrt{r_i} \ket{i}_{\CA} \ket{i}_{\CB}
$$
with local density matrix $\rho = \sum_i r_i \ket{i}\bra{i}$ on either side.
Upon performing a POVM on $\CA$, defined by the positive operators 
$\{ \Lambda_x \}$ with $\sum_i \Lambda_i = \1$, Alice
holds a random variable $X$ correlated with Bob's quantum system $\CB$.
Moreover, according to \cite{ensembles}, the ensemble of $X \CB$ is given by
$\{\rho_x, p(x) \}$, where 
\begin{eqnarray}
p(x)    & = &  \tr ( \rho \Lambda_x ), \nonumber\\
\rho_x  & = & \frac{1}{p(x)} [\sqrt{\rho} \Lambda_x \sqrt{\rho}]^* 
\label{pars}
\end{eqnarray}
and $*$ denotes complex conjugation in the $\{ \ket{i} \}$ basis.

A useful representation of classical-quantum systems, which
we refer to as the ``enlarged Hilbert space'' (EHS) representation,   
is obtained by embedding the random variable $X$ in some quantum system $\CA$.
Then our ensemble $\{ \rho_x, p(x) \}$ corresponds to the density 
operator 
\beq
\rho^{\CA \CQ} = \sum_x p(x) 
\ket{x}\bra{x}^{\CA} \otimes \rho_x^{\CQ},
\label{cq}
\eeq
where $\{\ket{x}: x \in \CX \}$ is an orthonormal basis for the Hilbert space 
$\CH_\CA$ of $\CA$. A classical-quantum system may, therefore, be viewed 
as a special case of a quantum one.
The EHS representation is convenient for
defining various information theoretical
quantities for classical-quantum systems. The von Neumann entropy of a quantum 
system $\CA$ with density operator 
$\rho^\CA$ is defined as $H(\CA) = - \tr \rho^\CA \log \rho^\CA$.
For a bipartite quantum system $\CA \CB$ define the conditional von 
Neumann entropy 
$$
H(\CB| \CA) = H(\CA \CB) - H(\CA),
$$
and quantum mutual information
$$
I(\CA; \CB) = H(\CA) + H(\CB) - H(\CA \CB) = H(\CB) - H(\CB| \CA),
$$
in formal analogy with the classical definitions.
Notice that for classical-quantum correlations ($\ref{cq}$)
the von Neumann entropy $H(\CA)$ is just the
Shannon entropy $H(X) = - \sum_x p(x) \log \, p(x)$ of $X$.  
The conditional entropy $H(\CQ|X)$ is defined as $H(\CQ|\CA)$ and
equals $\sum_x p(x) H(\rho_x)$.
Similarly, the mutual information of $X \CQ$ 
is defined as  $I(X; \CQ) = I(\CA; \CQ)$. Notice that this
is precisely the familiar Holevo information \cite{holold}
of the ensemble $\CE$:
$$\chi(\CE) = H\left( \sum_x p(x) \rho_x \right) - \sum_x p(x) H(\rho_x).
$$ 

Returning to the formulation of the CQSW problem, 
suppose Alice and Bob share a large number $n$ copies of
the classical-quantum system $X \CQ$. 
Alice possesses knowledge of the index $x^n = x_1 x_2 \dots x_n$, 
but not the quantum system locally described by  
$\rho_{x^n} = \rho_{x_1} \otimes \rho_{x_2} \cdots \otimes \rho_{x_n}$;
Bob has the quantum system at his disposal but not the classical index.
Note that this does not necessarily imply that Alice can prepare a replica
of Bob's state in a way that preserves its entanglement with other systems.
Alice wishes to convey the information contained in the
index $x^n$ to Bob almost perfectly, using a minimal amount of classical communication. 
If Bob didn't have the  quantum information, she would need to 
send $\approx n H(X)$ classical bits. 
The question is: can they reduce the communication cost
by making use of Bob's quantum information? 
To consider a trivial example, the members of the ensemble could be
mutually orthogonal. Then Bob would be able to perfectly distinguish among 
them by performing an appropriate measurement, requiring no classical communication
whatsoever. 
An intermediate case is when $X \CQ$ is given by the BB84 \cite{BB84} ensemble 
$\CE_{\rm{BB}84}$. Taking  $\{ \ket{0}, \ket{1} \}$ to be the standard qubit basis, 
let
$\ket{\pm} = \frac {1}{\sqrt{2}} (\ket{0} \pm \ket{1})$. 
$\CE_{\rm{BB}84}$ assigns a probability of 
$\frac{1}{4}$ to each of $\ket{0}, \ket{1} ,\ket{+}$ and $\ket{-}$, so that  $2$ bits 
are required to describe Alice's classical data. However, she 
needs to send only $1$ bit indicating the basis $\{ \ket{0}, \ket{1} \}$ 
or $\{ \ket{+}, \ket{-}\}$ in which Bob should perform his measurement.
The measurement unambiguously reveals the identity of the chosen 
state without disturbing it. This example is a one-shot paradigm for the general
case.  A single copy of a general $X \CQ$ does not have this
property of being decomposable into subensembles with mutually orthogonal 
elements. However, the block $X^n \CQ^n = X_1 \CQ_1 X_2 \CQ_2 \dots
X_n \CQ_n$, consisting of a large number of
copies of $X \CQ$, does satisfy this condition approximately. 
Since the problem is formulated as an asymptotic and approximate one, 
this will suffice for our purposes.
We shall show that Alice may reduce her communication cost by at most
$\approx n I(X; \CQ)$, and describe a protocol that achieves this.
We proceed to formally define the coding procedure.
An $(n, \epsilon)$ CQSW code consists of 

$\bullet$ a mapping $f: \CX^n \rightarrow [M]$,  $[M] = \{ 1,2, \dots, M \}$,  
$M = 2^ {n R}$, by which Alice encodes her classical message
$X^n$ into the index  $I = f(X^n)$;

$\bullet$ a set $\{ \Lambda^{(1)}, \Lambda^{(2)}, \dots, \Lambda^{(M)}  \}$, where
each $\Lambda^{(i)} = \{\Lambda^{(i)}_j  \} $ is a POVM
acting on $\CH^{\otimes n}$ and taking values $j \in [N]$;

$\bullet$ a decoding map $g: [M] \times [N] \rightarrow \CX^n$ 
that provides Bob with an estimate $\hat{X}^n = g(I, J)$ of $X^n$
based on $I$ and the outcome $J$ of the POVM $\Lambda^{(I)}$
applied to Bob's quantum system $\CQ^n$.

\vspace{2mm}

The \emph{rate} $R$ signifies the number of bits per copy needed to 
encode the index $I$. The error probability is required to 
be bounded 
$$
P_e = \Pr \{ \hat{X}^n \neq {X}^n \} \leq \epsilon.
$$
Denoting Bob's residual state after the
extraction of the classical information by 
$\hat{\rho}_{x^n}$, its disturbance with respect to
$\rho_{x^n}$ must also be small on average
\beq
\Delta = \sum_{x^n} p(x^n) \,
\| \hat{\rho}_{x^n} - \rho_{x^n} \|_1 \leq \epsilon.
\label{tact}
\eeq
A rate $R$ is said 
to be \emph{achievable} if for
any $\epsilon, \delta > 0$ and all sufficiently large $n$ 
there exists an $(n, \epsilon)$ code of rate $R + \delta$.
Our main result (which first appeared in \cite{winterphd})
is the following theorem.
\begin{thm1}\emph{(CQSW Theorem)}
Given a classical-quantum system $X \CQ$, 
a rate $R$ is achievable iff 
$$R \geq H(X) - I(X; Q) = H(X| \CQ).$$
\end{thm1}
\noindent 
The ``if'' part of the proof is called the direct coding theorem, and
the ``only if'' part is called the converse. 

\begin{figure}
\centerline{ {\scalebox{0.50}{\includegraphics{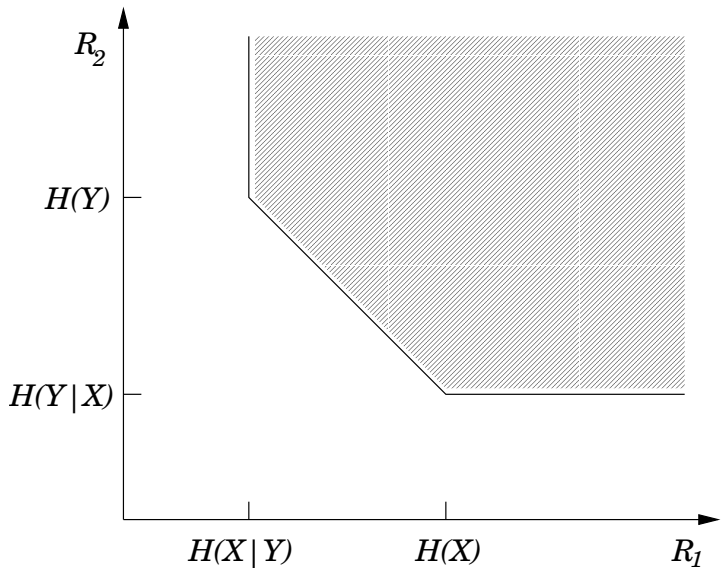}}}}

\caption{The achievable rate region for the classical Slepian-Wolf problem.}
\label{sw-fig}
\end{figure}

Let us first compare our result to the classical Slepian-Wolf problem.
The latter is  usually formulated as a three terminal problem.
We are given two correlated sources described by the random variables $X$ and $Y$,
known to  Alice and Bob, respectively.
They encode their sources separately and send them to Charlie at rates $R_1$ and 
$R_2$, respectively, who  decodes them 
jointly with the aim to faithfully reconstruct $X$ and $Y$.
One may now ask about the achievable rate \emph{region} $(R_1, R_2)$. 
The answer is given by
\begin{eqnarray*} 
R_1 &\geq & H(X|Y) \\
R_2 &\geq&  H(Y|X) \\
R_1 + R_2 &\geq&  H(XY),
\end{eqnarray*} 
as shown in figure 1. 
It suffices to show the achievability of the points 
$(H(X), H(Y|X))$ and  $(H(X|Y), H(Y))$, since the rest of the region follows
by time sharing (merging codes of length $\nu n$ and $(1 - \nu) n$, 
$ \nu \in [0,1]$, corresponding the two points, respectively).
The obvious classical-quantum generalization of this result would be 
to replace $Y$ by a quantum system $\CQ$, and the joint distribution of $XY$
by the ensemble state
\beq
\rho^{\CA\CQ} = \sum_{x,y} p(x,y)\ket{x}\bra{x}\otimes\pi_y,
\label{cq-source}
\eeq
where $\pi_y$ are density operators on $\CQ$, which for the sake of this
discussion we assume to be pure. Observe that the state written here has
the same form as in (\ref{cq}), with
\begin{eqnarray}
  p(x)   &=& \sum_y p(x,y), \nonumber \\
  \rho_x &=& \frac{1}{p(x)}\sum_y p(x,y)\pi_y. \nonumber
\end{eqnarray}
But here the description also contains the decomposition of $\rho_x$ into
pure states, i.e., a chosen ensemble.

The task of coding is, analogously to Schumacher's theorem, to enable Charlie
to reconstruct $\ket{x^n}\bra{x^n}\otimes\pi_{y^n}$ with high average fidelity,
in a situation of many independent realizations of $\rho^{\CA\CQ}$.
Indeed, Theorem 1 implies the achievability of the point $(H(X|\CQ), H(\CQ))$. 
Bob may Schumacher compress his quantum system and send it to 
Charlie at a qubit rate of $R_2 = H(\CQ)$.
The latter uses it as quantum side information, and 
Alice needs to send classical information to Charlie at a 
bit rate of $R_1 = H(X|\CQ)$. 
Furthermore, after having used the quantum system for this purpose,
according to (\ref{tact}) it will remain basically intact. (Note that
our proof of the direct coding theorem below actually shows that
even the average disturbance of the $\pi_{y^n}$ is small -- in fact
the decoder is such that it causes little disturbance to
\emph{purifications of the $\rho_{x^n}$}.)

As for the other point $(H(X), H(\CQ|X))$, we do not know to which extent
the classical result carries over.
There are, as in the above discussion, trivial examples where
it is achievable. One example is perfect correlation, when $p(x,y)\neq 0$
iff $x=y$: then, knowing $x$ one can perfectly reconstruct the pure
state of $\CQ$ because it has to be $\pi_x$. So, $R_1=H(X)$, $R_2=0$
is achievable.
Another is when $X$ can be read off $\CQ$, i.e. when the states $\pi_y$
fall into mutually orthogonal classes $\CY_x$ such that $p(x,y)\neq 0$
implies $\pi_y\in\CY_x$. Then Alice can Shannon compress her $x^n$,
and Bob, since he can read $x^n$ on his system, can Schumacher
compress to a rate $H(\CQ|X)$ (compare\cite{devberger} and \cite{HJW}).

Notice that there are two variants to the coding problem here:
\emph{blind} (where Bob has to operate on the $\pi_y$), and \emph{visible}
(where he is told $y$). Note that the labeling of the different ensembles
for the $\rho_x$ by the same set $\CY$ is purely artificial -- this is why
there is more than one visible coding problem associated to the same
ensemble. In particular, we cannot expect the answers to the visible and
to the blind problem to be the same. Both however are open problems.

\vspace{3mm}

\noindent {\bf{Proof of Theorem 1 (converse)}}    \space \space 
We need to prove that, for any $\delta, \epsilon > 0$ and 
sufficiently large $n$, if an $(n, \epsilon)$ code has rate
$R$  then $R \geq H(X|\CQ) - \delta$. 
Without loss of generality,  $\epsilon \leq \frac{\delta}{2 |\CX|}$ 
and $n \geq 2/\delta$. We shall make use of two
inequalities. The first is the Holevo bound \cite{holold}, according
to which the amount of information about $X^n$ extractable from the
quantum system $\CQ^n$ is bounded from above by $I(X^n; \CQ^n) = n I(X; \CQ)$.
Recall that Bob makes an estimate  $\hat{X}^n = g(I,J)$ of $X^n$ based
on $I = f(X^n)$ and the measurement outcome $J$.
Our second ingredient is  Fano's inequality \cite{ct}:
$$H(X^n| I J) \leq h_2(P_e) + P_e \log (|\CX|^n  - 1).$$
Here $h_2(p) = -p \log p - (1 - p) \log (1 - p)$ is the binary entropy.
This inequality is interpreted as: Given $I J$ one can specify
$X^n$ by saying whether or not it is equal to $g(I,J)$ and, conditionally
upon a negative answer, specifying which of the remaining $|\CX|^n  - 1$ 
values it has taken. We have
\begin{eqnarray}
\lefteqn{n R + n I(X; \CQ)}   \nonumber\\
& \geq & H(I) + I(X^n; J) \nonumber\\
&  = & H(X^n) + H(I| X^n J) + I( I; J) - H(X^n| I J) \nonumber\\
& \geq & n H(X)  - H(X^n| I J) \nonumber\\
& \geq & n \left( H(X)  - \frac{1}{n} - \epsilon \log |\CX| \right). \nonumber
\end{eqnarray}
The first inequality follows trivially from  $I \in [ 2^{n R}] $
and  Holevo's theorem. The second comes from
the non-negativity of mutual information and conditional entropy.
The final one is a consequence of Fano's inequality.
Thus $R \geq H(X) - I(X; \CQ) - \delta$, as claimed. \qed

\noindent {\bf Remark} \,\,  An alternative way to demonstrate the converse 
uses a recent result 
on remote state preparation \cite{newRSP}, according to which Alice and
Bob may establish classical-quantum correlations $X \CQ$ with asymptotically 
perfect fidelity using shared entanglement and forward classical communication 
at rates of $H(\CQ)$ ebits and $I(X; \CQ)$ bits, respectively.
Let us assume that the converse fails, i.e. that  it is possible to achieve a 
CQSW rate  $R < H(X|\CQ)$. 
Then with the help of shared entanglement she would be able 
to convey $X$ at a classical rate strictly less than $H(X)$, by first 
remotely preparing the quantum information then performing the CQSW protocol. 
We know, however, that entanglement can in no way increase the capacity of a 
classical channel, e.g. by \cite{eac}.

\noindent {\bf Remark} \,\,  Note that the lower bound in Theorem 1 holds
true even for CQSW codes which disregard condition (\ref{tact}): We invite
the reader to confirm that in the proof of the converse it was never used.

\vspace{2mm}

Before launching into the proof of achievability we give a heuristic
argument. Let us recall typical sequences (see \cite{ck} for an extensive
discussion) and subspaces \cite{nono} and their properties. 
The theorem of typical sequences states that given  random variable $X$ defined
on a set $\CX$ and with probability distribution $p(x)$, 
for any $\epsilon, \delta > 0$ and  sufficiently large 
$n > n_0(|{\cal X}|, \epsilon, \delta)$ 
there exists a \emph{typical set} $T_{X, \delta} \subset \CX^n$ of sequences 
$x^n$ such that $$2^{n [H(X) - \de]} \leq |T_{X,\de}|  \leq 2^{n [H(X) + \de]},$$
and $\Pr\{X^n \in T_{X, \delta} \} \geq 1 - \epsilon$.   
Typical sequences are those in which the fraction of a given
letter $x$ is approximated by its probability $p(x)$, and the law of large
numbers guarantees that such sequences will occur with high probability.
Thus one need worry only about encoding typical sequences.
The quantum analogue of the typical set is the \emph{typical subspace}
$\CT_{\CQ, \delta}$ of $\CH^{\otimes n}$,
defined for a quantum system $\CQ$ with $d$-dimensional Hilbert
space $\CH$ and in a quantum state $\rho$. It satisfies
$$2^{n [H(\CQ) - \de]} \leq \dim \CT_{\CQ, \delta}  \leq 2^{n [H(\CQ) + \de]},$$
and $\tr (\rho^{\otimes n} \Pi_{\CQ, \delta})  \geq 1 - \epsilon$, where
$ \Pi_{\CQ, \delta}$ is the projector onto $\CT_{\CQ, \delta}$.
Finally, for a classical quantum system $X \CQ$ 
and a particular sequence $x^n \in T_{X,\de}$ we define the  
\emph{conditionally typical subspace} $\CT_{\CQ|X, \delta}(x^n)$ in the following way.
The Hilbert space $\CH^{\otimes n}$ can be decomposed
into a tensor product $\bigotimes_{x} \CH_x$ with $\CH_x$ collecting all the factors
$k$ such that $x_k = x$. Then the conditionally typical subspace is the tensor
product of the typical subspaces of the $\CH_x$ with respect to 
${\rho}_x$. It follows that
$$2^{n [H(\CQ|X) - K \de]} \leq \dim \CT_{\CQ|X, \delta}(x^n)
 \leq 2^{n [H(\CQ|X) + K \de]},$$
for some constant $K$.
At the same time $\tr (\rho_{{x^n}} \Pi_{\CQ|X, \delta}(x^n))  > 1 - |\CX| \epsilon$, 
where $\Pi_{\CQ|X, \delta}(x^n)$ is the projector onto $\CT_{\CQ|X, \delta}(x^n)$.
The latter means that the trace decreasing  measurement given by
$\Pi_{\CQ|X, \delta}(x^n)$ will succeed with high probability when applied
to the state $\rho_{{x^n}}$. 
One would like to construct a POVM out
of such conditionally typical projectors for different ${x^n}$ belonging
to some set $\CC$, in order to distinguish between them. 
Since the $\CT_{\CQ|X, \delta}(x^n)$ are approximately contained
in $\CT_{\CQ, \delta}$ \cite{strong},
the task is, roughly speaking, to ``pack''  the 
$\CT_{\CQ|X, \delta}(x^n)$, ${x^n} \in \CC$ 
into the typical subspace $\CT_{\CQ, \delta}$. The former have dimension 
$\doteq 2^{n H(\CQ|X)}$ and the latter has dimension $\doteq 2^{n H(\CQ)}$,
hence one expects $|\CC|$ to be at most 
$\doteq 2^{n [H(\CQ) - H(\CQ|X)]} = 2^{n I(X;\CQ)}$.
This is the basic content of the HSW, or $\{c \rightarrow q\}$
channel coding, theorem \cite{holevo},
although the actual POVM construction is rather more subtle. 
Accordingly, $\CC$ is called a \emph{channel code}.
Here we take one step further and ask about the minimal
number of disjoint channel codes that ``cover'' the typical input set 
$T_{X, \delta}$. The size of $T_{X, \delta}$ is $\doteq 2^{n H(X)}$, so
the number of codes needed should be $\doteq 2^{n [H(X) - I(X;\CQ)]}$. 
Now Alice need only send information about which code her source sequence 
$x^n$ belongs to, 
and Bob can perform the appropriate measurement to distinguish it from the other
sequences in that code, as in the one-shot BB84 example. The described construction
is depicted in figure 2.

\begin{figure}
\centerline{ {\scalebox{.53}{\includegraphics{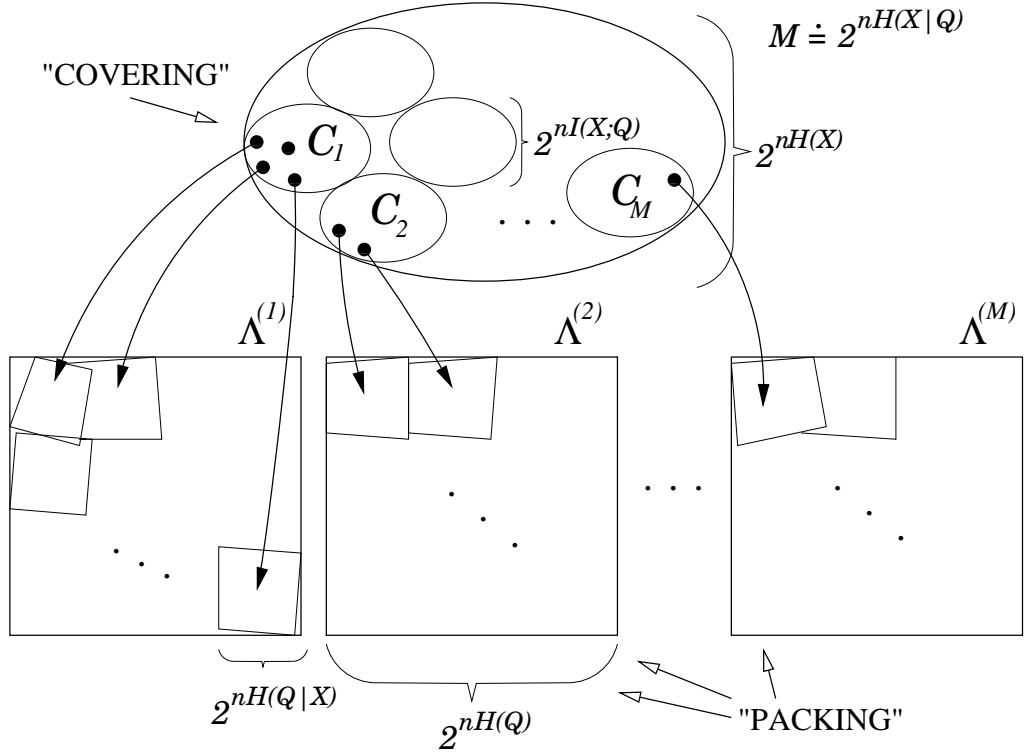}}}}

\caption{A simple counting argument for the optimal CQSW rate.}

\end{figure}

To prove Theorem 1 we shall need some background on channel codes. 
For a given classical-quantum system $X\CQ$, 
a channel code $\CC$ is a subset 
of $\CX^n$, associated with a POVM $\Lambda = \{ \Lambda_{x^n}: {x^n} \in \CC \}$ 
acting on $\CH^{\otimes n}$.
The \emph{rate} of the channel code is defined as $r = \frac{1}{n} \log |\CC|$.
The error probability of a given ${x^n} \in \CC$ is  
$p_{e}({x^n}) = 1 - \tr (\rho_{{x^n}} \Lambda_{{x^n}})$. 
$\CC$ is called an $(n, \epsilon)$ code if 
$\max_{{x^n} \in \CC} p_{e}({x^n}) \leq \epsilon$.
We shall need the following version of the 
$\{c \rightarrow q\}$ channel coding theorem \cite{strong}:

\begin{thm2}\emph{(Winter \cite{strong}, Theorem 10)}
For all $\eta,  {\epsilon},{\de} \in (0,1)$, sufficiently large
$n \geq  n_1(|\CX|,d,\eta, {\epsilon},{\de})$
and every subset $\CA \in \CX^n$ with $\Pr \{{x^n} \in \CA\} \geq \eta$,
there exists an $(n, {\epsilon})$ channel code $\CC$ 
of rate $r \geq I(X; \CQ) - {\de}$ satisfying $\CC \subset \CA$. 
\end{thm2}
\noindent 
The $\CC \subset \CA$ condition is sufficiently strong to easily yield
the achievability part of the CQSW theorem,
following a standard classical argument of Csisz\'{a}r and  K\"orner \cite{ck}. 
 
\vspace{2mm}

\noindent {\bf{Proof of Theorem 1 (coding)}}    \space \space 
Fixing $0 < {\epsilon} < \frac{1}{2}$ and ${\de} > 0$
we shall first show that for sufficiently large $n$ there exists a family of 
disjoint channel codes $\{ \CC_1, \CC_2, \dots , \CC_{M-1} \}$  such that 
$$\Pr ({x^n} \notin \bigcup_{i = 1}^{M-1} \CC_i) \leq 2 {\epsilon}$$
and $\frac{1}{n} \log M \leq H(X|\CQ) + 2 \de$, thus 
upper bounding the number of channel codes needed to cover most of the high 
probability sequences. Recall that
for $n \geq n_0(|{\cal X}|, {\epsilon}, \de)$  we have 
$\Pr(X^n \in T_{{\de}}(X)) > 1 - {\epsilon}$. 
By Theorem 2 we also have that for
$n \geq n_1(|{\cal X}|,d,{\eta},{\epsilon},{\de})$
and every subset $\CA \in \CX^n$ with $\Pr \{{x^n} \in \CA\} \geq {\epsilon}$
there exists an $(n, {\epsilon})$ code of rate $r \geq I(X; \CQ) -  {\de}$
satisfying  $\CC \subset \CA$. We choose $n \geq \max \{ n_0,n_1 \}$
so that both conditions are satisfied. The idea is to keep constructing disjoint 
codes from $T_{X,\de}$ for as long as  Theorem 2 allows.
Define $\CA_1 = T_{X,\de}$, and let $\CC_1 \subset \CA_1$ be an 
$(n, {\epsilon})$
code as specified by Theorem 2. Recursively construct in a similar manner 
 $\CC_i \subset \CA_i$ where $\CA_i = T_{X,\de} - \bigcup_{j = 1}^{i} \CC_j$, 
which will also satisfy the conditions of the theorem
as long as $\Pr \{{x^n} \in \CA_i\} \geq {\epsilon}$. Suppose the construction
stops at $i = M$, i.e. $\Pr \{{x^n} \in \CA_M\} \leq  {\epsilon}$ . Then we have 
\beq
\Pr \{{x^n} \notin \bigcup_{i = 1}^{M-1} \CC_i\} = 
\Pr\{X^n \notin T_{X,\de}\}
+  \Pr \{{x^n} \in \CA_M\} \leq 2 {\epsilon}.
\label{eq1}
\eeq
On the other hand 
$$ 2^  {n [H(X) +{\de}]} \geq  
|T_{X,\de}| \geq \sum_{i = 1}^{M-1} |\CC_i| \geq (M-1) \, 2^{n[I(X;\CQ)- {\de}]}, $$
which implies
$$
R = \frac{1}{n} \log M \leq H(X) - I(X;\CQ) + 2 \de.
$$
The mapping $f$ is now defined as
$$
f({x^n}) =  \left\{ \begin{array}{ll}
i & { {x^n} \in \CC_i}\\
M & {\rm{otherwise}} 
\end{array} \right.
$$
The latter case, which signifies an encoding error,
happens with probability $ \leq 2 {\epsilon}$ by (\ref{eq1}).
Otherwise, Bob performs the POVM corresponding to the code
$\CC_{f(x^n)}$, which fails to correctly identify $x^n$ with
probability $\leq \epsilon$.
Therefore the total error probability is bounded $P_e \leq  3 \epsilon $.
Finally, Winter's ``gentle measurement'' lemma \cite{mac}, 
which states that a POVM with a highly predictable outcome 
on a given state cannot disturb it much,
guarantees that the average disturbance $\Delta$ is bounded 
by $\sqrt{8 \epsilon} + \epsilon$. 
The direct coding theorem follows. \qed
 
\noindent {\bf Remark} \,\,  The ``gentle measurement'' lemma invoked
at the end of the proof actually applies equally if the measurement
acts on one half of a purification of the state $\rho_{x^n}$ -- a fact
we needed in the discussion of the Slepian-Wolf theorem after the statement
of Theorem 1.

\vspace{2mm}

Finally, we would like to comment on a connection to Winter's measurement 
compression theorem \cite{winter}. 
Suppose Bob needs to perform a ``BB84'' measurement  given
by the operation elements $\{ \frac{1}{2} \ket{0} \bra{0}, \frac{1}{2} \ket{1} \bra{1},
\frac{1}{2} \ket{+} \bra{+}, \frac{1}{2} \ket{-} \bra{-} \}$ on a quantum system described
by the uniform density matrix.
He would then need $2$ classical bits of communication to convey the 
outcome to Alice. 
Equivalently, he can use $1$ bit of shared randomness between
him and Alice to decide which of the two measurements 
$\{ \ket{0} \bra{0}, \ket{1} \bra{1} \}$ or $\{ \ket{+} \bra{+}, \ket{-} \bra{-} \}$
he should perform, and send her only $1$ bit describing the outcome.
He has thus perfectly \emph{simulated} the measurement, 
replacing $1$ bit of communication with the weaker resource of
$1$ bit of shared randomness. 

For a general source-POVM pair $(\rho, \Lambda = \{\Lambda_x\} )$,
define the classical system $X \CQ$ by the ensemble $\{ \rho_x, p(x)\}$,
given by (\ref{pars}); in other words $X \CQ$ embodies the correlations
between the measurement outcome $X$ (to be sent to Alice)
and the reference system $\CQ$ that purifies the system to be measured. 
In an asymptotic and approximate setting,
the measurement $\Lambda^{\otimes n}$ is considered 
well simulated on $\rho^{\otimes n}$ 
if the classical-quantum correlations established
between Alice and Bob's reference system closely resemble $n$ copies of $X \CQ$. 
It was shown in \cite{winter} that the optimal classical communication 
and shared randomness rates become $I(X; \CQ)$ and $H(X| \CQ)$ respectively.
It is not surprising that the minimal amount of classical communication
required to establish a remote classical-quantum correlation is given by
the corresponding Holevo information.
Achievability of this bound may described by a diagram similar to the one 
depicted in figure 2, with 
the difference that ``PACKING'' should be replaced by ``COVERING''.
The idea is to divide the set of typical outcome sequences $T_{X, \delta}$ into 
codes $\CC_i, i \in [M]$, such that $\{\rho_{x^n}: x^n \in \CC_i \}$
mimics the set of residual states of the reference  system 
after performing some measurement  $\Lambda^{(i)}$. 
Thus $|C_i|$ must be sufficiently large to allow
$$
\sum_{x^n \in \CC_i} p(x^n) \rho_{x^n} \approx const. \times  \rho^{\otimes n},  
\,\,\, \forall i \in [M].
$$
Since $\rho^{\otimes n}$ and $\rho_{x^n}$ are ``almost'' uniformly
supported on $\CT_{\CQ, \delta}$ and $\CT_{\CQ|X, \delta}(x^n)$, respectively,
\cite{strong}, dimension counting arguments again suggest 
$|\CC_i| \doteq 2^{n I(X;\CQ)}$; moreover  $M \doteq 2^{n I(X|\CQ)}$ as before.
$\Lambda^{\otimes n}$ is then simulated by randomly choosing one of the
$\Lambda^{(i)}$, as in the BB84 example.

\vspace{2mm}

Coding with side information is a relatively unexplored and potentially rich area of 
quantum information theory. We have presented here an important member of
this class of problems, providing yet another example of classical Shannon 
theory generalizing to the quantum domain.

Our main result can be understood as the translation of one of the two extremal
rate points of the classical Slepian-Wolf region to a classical-quantum
scenario.
The main question left open is whether one can also translate the other rate point;
it might actually be that the rate region in the quantum case does not
look like figure \ref{sw-fig}.

\vspace{2mm}

\noindent {\bf{Acknowledgments}}\,\,
We are indebted to Debbie Leung for comments on an earlier version of the manuscript.
Thanks also go to Charles Bennett, Toby Berger, Andrew Childs, Patrick Hayden,
Aram Harrow, Luis Lastras, Anthony Ndirango and John Smolin for useful discussions. 

ID is supported in part by the NSA under the US Army Research Office (ARO), 
grant numbers DAAG55-98-C-0041 and DAAD19-01-1-06. 
AW is supported by the U.K.~Engineering and Physical Sciences Research Council.

\end{document}